# Chirality tuning and reversing with resonant phase-change metasurfaces

Xinbo Sha[1,2], Kang Du[1], Yixuan Zeng[1], Fangxing Lai[1], Jun Yin[1], Hanxu Zhang[3], Bo Song[3], Jiecai Han[3], Shumin Xiao[1,2,3,#], Yuri Kivshar[4,§], Qinghai Song[1,2,5,*]

1. Ministry of Industry and Information Technology Key Lab of Micro-Nano Optoelectronic Information System, Guangdong Provincial Key Laboratory of Semiconductor Optoelectronic Materials and Intelligent Photonic Systems, Harbin Institute of Technology, Shenzhen 518055, P. R. China.
2. Pengcheng Laboratory, Shenzhen 518055, P. R. China.
3. National Key Laboratory of Science and Technology on Advanced Composites in Special Environments, Harbin Institute of Technology, Harbin 150080, P. R. China.
4. Nonlinear Physics Center, Research School of Physics, Australian National University, Canberra ACT 2601, Australia.
5. Collaborative Innovation Center of Extreme Optics, Shanxi University, Taiyuan 030006, Shanxi, P. R. China

## Abstract

Dynamic control of circular dichroism in photonic structures is critically important for compact spectrometers, stereoscopic displays, and information processing exploiting multiple degrees of freedom. Metasurfaces can help miniaturize chiral devices but only produce static and limited chiral responses. While external stimuli are able to tune resonances, their modulations are often weak, and reversing continuously the sign of circular dichroism is extremely challenging. Here, we demonstrate dynamically tunable chiral response of resonant metasurfaces supporting chiral bound states in the continuum combining them with phase-change materials. Phase transition between amorphous and crystalline phases allows to control chiral response and vary chirality rapidly from −0.947 to +0.958 backward and forward via chirality continuum. Our demonstrations underpin the rapid development of chiral photonics and its applications.

Chirality refers to the property of an object or system that cannot be mapped to its mirror image by rotations and translations. [1] It is ubiquitous in all levels of nature being closely related to the origin of life. Chiroptic response of natural media is usually very weak and static, so artificial structures were suggested to enhance and tailor the optical chirality. [2, 3] Intrinsic or extrinsic circular dichroism (CD) has significantly been improved by top-down fabricated and bottom-up synthesized nanostructures during the past two decades. [4-15] Recently, the concept of bound states in the continuum (BICs) [16-17] has been employed to maximize chiroptical effects. By controlling in-plane and out-of-plane asymmetry, maximized intrinsic chirality and high-quality factor (Q factor) can be achieved with resonant metasurfaces realizing the concept of *chiral BICs*. [18-20] Consequently, chiral light-matter interactions have been significantly enhanced, and various novel phenomena, such as chiral spontaneous emission, chiral lasing, and chiral mode splitting, have been proposed and experimentally demonstrated in nanophotonic metadevices. [21-24] Despite ongoing breakthroughs, current demonstrations yield only static and limited chiroptical responses, which hardly meet the needs of applications including chiral detection, stereoscopic displays, and information processing.

Being different for molecules, chiroptical effects in nanophotonics arise from the interactions between structured materials and internal electromagnetic fields. [4] In this sense, optical response can be tailored by reconfiguring nanostructures and controlling their internal field distributions with an external stimulus. [2] Reconfigurable chiroptical response has intensively been explored in microwave and terahertz fields. [25-27] Chirality of designed metadevices is controllable under optical or mechanical stimuli. [25-27] With the assistance of phase-change materials (PCM), liquid crystals, DNA origami, NEMS, and magnetic fields, reconfigurable chiroptical effects have also recently been achieved in optics. [28-35] Nevertheless, most of the aforementioned tunable chirality in optical metadevices is merely switched between on and off. [31-37] Reversing handedness is typically a binary effect restricted to very small CD values. [28-30] Up to now, the strategy for reversing large CD values and realizing a continuous change of the CD values (the so-called chirality continuum [38]) is still absent. Here, we combine the concept of chiral quasi-BICs with PCMs to address this long-standing challenge and demonstrate hybrid metasurfaces with transformational and tunable chiral response.

Figure 1(a) shows schematically our concept. A metasurface is created by a square-lattice periodic structure with lattice spacings of $p_x$ = 975 nm and $p_y$ = 905 nm. Each unit contains two tilted Si nanobars with in-plane asymmetry $\theta$ and out-of-plane asymmetry $\alpha$. The length, width, separation distance, and height of Si nanobars are $l$ = 700 nm, $w$ = 220 nm, $d$ = 460 nm, and $h$ = 250 nm, respectively. The Si bars are covered with 490 nm Polyimide, 30 nm layer of PCM (Antimony trisulfide $Sb_2S_3$), and 15 nm of $SiO_2$. [39-44] Optical properties of such a hybrid metasurface are studied numerically (see Methods) with parameters taken from the experimental results (Supplementary Note-1). The bandgap structure for amorphous-phase metasurface with $\theta$ = 15° and $\alpha$ = 0° is illustrated in Fig. 1(b). Three modes can be observed there, and two of them are associated with quasi-BICs and high Q factors (Supplementary Note-1). Such resonances do not possess optical chirality since mirror-reflection symmetry of the metasurface without a substrate is preserved in the vertical direction.

The mirror-reflection symmetry can be broken by introducing a vertical tilt with the angle $\alpha$. The slanted Si bar pairs redistribute their fields and induce an offset ($d$) between two effective dipoles. The coupling parameters of LCP and RCP waves ($m_\pm$) to a metasurface are proportional to $\sin(kd/2 \mp \theta)$. The offset makes different the coupling parameters and produces a chiral response (Supplementary Note-2). Figure 1(c) shows the dependence of the effective vertical offsets ($d$) on the tilt angle for two transverse magnetic (TM) polarized modes. While these two modes have similar wavelengths, their response to the tilt angle is completely different. The offset of one mode grows monotonically from zero as the tilt angle increases. The other mode, however, gives a negative value even though it doesn't change linearly due to the mode interaction (Supplementary Note-2). This trend can also be observed in Fig. 1(d) from the field patterns ($|H_{in\text{-}plane}|$), where the positions of maximal fields of two modes are reversed.

For a fixed in-plane deformation parameter $\theta$, the opposite signs of offsets of two resonances (Fig. 1(c)) produce opposite chiral responses ($m_\pm \propto \sin(kd/2 \mp \theta)$). Thus, it is possible to find a set of parameters $\theta$ and $\alpha$ to realize two modes with maximized and opposite optical chirality, i.e., chiral quasi-BICs with $CD = \pm 1$. One example is depicted in Fig. 1(e). When the in-plane and out-of-plane asymmetries are $\theta = 15º$ and $\alpha = 20º$, two resonances at 1499 nm and 1505 nm can be seen in the RCP and LCP transmission spectra, respectively. Following the definition $CD = (T_{LCP} - T_{RCP})/(T_{LCP} + T_{RCP})$, we calculate optical chirality of phase-change metasurface under amorphous phase and show it by a solid line in Fig. 1(g). The CD values of two resonances are very close to ±1, respectively. Then we know that the same metasurface is able to simultaneously produce two opposite chiral quasi-BICs at the Γ-point.

To introduce tunability, we cover Si metasurface with a PCM layer of 30 nm. A change in refractive index of PCM during phase transition [39-44] perturbs only the resonant wavelengths instead of significantly modifying the main field patterns and corresponding chiral response. Figure 1(f) shows the LCP and RCP transmission spectra of crystalline-phase metasurface. Two chiral resonances can still be observed except that their wavelengths shift to 1505 nm and 1511 nm, respectively. The corresponding CD value has also been calculated and plotted as a dashed line in Fig. 1(g). We can see that the optical chirality is kept at CD = 0.944 and CD = -0.981, which are close to the values of amorphous phase metasurface well. One interesting phenomenon rises at the spectral range around 1505 nm (shadowed region in Fig. 1(g)). The CD value changes from -0.979 to 0.944 with the transition from amorphous to crystalline phase. These results are consistent with the above analysis, and they suggest that phase-change metasurfaces can be employed for the realization of dynamically controllable optical chirality.

Based on the above analysis and numerical simulation, we fabricate the designed Si nano-bars with a combined process of electron-beam evaporation, electron-beam lithography, and reactive ion etching (Supplementary Note-3). [45] Figure 2(a) displays the high-resolution top-view scanning electron microscope (SEM) image of Si nano-bars. All the structural parameters including length, width, separation distance, and in-plane asymmetry $\theta$ follow the numerical design very well. Then the Si nano-bars are re-filled with Polyimide and further coated with $Sb_2S_3$ via thermal evaporation (see Methods and Supplementary Note-3). [46, 47] The entire metasurface is further coated with 15 nm $SiO_2$ as a protection layer. The corresponding side-view SEM image is depicted in Fig. 2(b), where the pillar height, tilt angle $\alpha$, and thickness of films are close to those designed numerically.

The hybrid metasurface is characterized by a home-made microscope system (Supplementary Note-4). Figures 2(c, e) show the angle-dependent transmission spectra under the illumination of LCP and RCP light. Both $TM_1$ and $TM_2$ modes can be observed, and their angular dependences match the numerical simulation well. We can see that $TM_1$ mode disappears at Γ-point in the RCP band structure (Fig. 2(e)), whereas $TM_2$ mode disappears at Γ-point in the LCP band (Fig. 2(c)). As a consequence, we confirm that two chiral quasi-BICs with opposite CD values have been simultaneously formed at 1501 nm and 1511 nm (see Fig. 2(g)).

Then the phase-changed metasurface is pumped optically by a 450 nm continuous wave (CW) laser with a pumping density of 41.7W/mm$^2$. Under such an external stimulus, PCM switches from amorphous phase to crystalline phase and thus its refractive index changes (see Supplementary Note-1). Figures 2(d, f) are the corresponding LCP and RCP angle-dependent transmission spectra of crystalline-phase metasurface. Similar to the amorphous phase, there are also two LCP and RCP resonances that disappear at Γ-point. Thus, two chiral quasi-BICs with opposite CD can still be observed (Fig. 2(h)). The only difference is their wavelengths shift to 1511 nm and 1519.8 nm. Then we know the dynamic reversal of optical chirality occurs at the spectral range around 1511 nm, consistent with theoretical model well.

The phase-change metasurface has very good device repeatability. We have fabricated a series of metasurfaces and characterized their optical response. The resonant wavelengths vary a few nanometers due to fabrication deviations. However, the designed CD values are well reproduced and maintained at a high level. Figure 3(a) shows the experimentally recorded CD spectra at Γ-point under amorphous phase and crystalline phase of one metasurface. For the spectral range around 1503.4 nm (shadowed area), a dramatic transition from CD = -0.954 to CD = 0.924 can be observed, consistent with the numerical simulation well. Figure 3(b) summarizes optical chirality of 12 samples under amorphous and crystalline phases (Supplementary Note-5). The average CD values before and after crystallization are around -0.9 and 0.85 for specific wavelengths. Here the CD values are larger than all previously reported for chiral quasi-BICs. The range of optical chirality inversion puts a new record for nanostructures and orders of magnitude higher than conventional materials. It can be further improved to CD = ±1 by finely adjusting the structural parameters (Supplementary Note-1).

For the case of tunable chirality, the reversibility is a critical characteristic. [46, 47] Under the excitation of a frequency doubled femtosecond laser (400 nm, 1kHz repetition rate, 100 fs pulse width, pumping density of 20 mJ/cm$^2$, Supplementary Note-4), PCM experiences a backward transition from crystalline to amorphous phase. In this sense, the performance of phase-change metasurface is reversible and its chiral response can be recovered to its original value. One example is shown in Fig. 3(c). We can see that the optical chirality in the shadowed area has been switched back and forth over multiple cycles. Figure 3(d) summarizes the experimental transition time of phase change process (Supplementary Note-4). Basically, the amorphous to crystalline transition is an annealing process requiring temperature above the glass transition temperature but below the melting point. Thus, the transition from amorphous to crystalline phase is relatively slow (bottom panel). In contrast, crystalline to amorphous transition requires a combined process melting and short annealing. The inverse transition time is as short as 800 ns (top panel). Both values are consistent with the published data [48, 49]

One intriguing property of $Sb_2S_3$ is the existence of multi-level intermediate states (see Fig. 4(a)). [46] Intermediate state is a mixture of crystalline phase and amorphous phase, and thus has an intermediate refractive index (Supplementary Note-6). The multi-level intermediate states of $Sb_2S_3$ trigger the possibility of continuously tunning the resonance of metasurface. Figure 4(b) shows the transmission spectra of a phase-change metasurface under different pumping density. When the power density (P) is below 25.83 W/mm$^2$, the phase-change metasurface remains unchanged. The LCP and RCP resonances appear at 1518 nm and 1510 nm, respectively. For the case of P > 25.83 W/mm$^2$, both resonances gradually shift to longer wavelength as a function of pumping power. The phase-change metasurface completely turns to crystalline state and the resonant wavelengths are fixed when the power exceeds 41.7 W/mm$^2$. For the spectral range around 1518 nm, the transmission of LCP light gradually increases, whereas the RCP transmission decreases. Then it is easy to know that the corresponding chirality can be changed continuously. The corresponding optical chirality is then calculated and plotted as dots in Fig. 4(c), where a gradual transition of optical chirality from CD = -0.947 to CD = 0.958 can be clearly observed.

In our experiment, the thickness of $Sb_2S_3$ is only 30 nm and thus the intermediate states can be considered as a uniform layer with different refractive indices. The corresponding refractive indices are obtained by fitting the transmission spectra and plotted as dots (#1, 6) in Fig. 4(d). These dots determine a straight line, which matches the other dots (#2- 5) well and reveals the linear dependence of refractive index on excitation power. Taking the refractive index on the line into account, we have numerically simulated the transmission spectra and calculated their corresponding CD values at all excitation power (Supplementary Note-6). As the dashed line shown in Fig. 4(c), the numerical results are consistent with the experimental observations very well. All intermediate states of PCM and the corresponding characteristics of phase change metasurfaces are found to be stable. Therefore, we can conclude that the chiral response of metasurface can be continuously tuned and the first stable chirality continuum covering CD = -0.947 to CD = 0.958 has been achieved experimentally.

In summary, we have combined the concepts of chiral quasi-BICs and phase-change materials to demonstrate hybrid resonant metasurfaces with the major functionalities to tune and reverse chiroptic response. By utilizing the inverse dependence of out-of-plane offset on asymmetry, we have achieved two resonances with maximum opposite chirality at the Γ-point of the spectrum. We have demonstrated that optical chirality of the resonant metasurface can be tuned continuously and reversibly between two extreme values via the transition between amorphous and crystalline PCM phases, thus realizing chirality continuum. Our demonstrations suggest a new paradigm for chiral nanophotonics underpinning many practical applications such as dynamically controlled chiral emission, chiral detection, and chiral displays. Performance and potential applications can be boosted further by employing electric tunning at the pixel level.

## METHODS

### Numerical simulations

Both metasurface transmittance and its eigenfrequencies are analyzed with the finite-element numerical solver in COMSOL Multiphysics. Periodic boundary conditions are applied in the x and y directions to mimic the in-plane infinite size of metasurface, whereas perfectly matched layers are utilized in the z direction to absorb the outgoing waves. The dielectric constants of all layers are taken from the experimental results measured by ellipsometry.

### Sample fabrication

The sample is prepared using standard nanofabrication technology. Amorphous silicon is deposited on a K9 glass substrate with electron beam evaporation (SKE_A_75). Then PMMA A2 electron-beam (E-beam) resist is spin-coated the silicon film with a speed of 4000 rpm. The designed nanostructures are patterned in PMMA film with E-beam lithography (Raith, e-line plus) and then developed in MIBK: IPA=1:3 for 30s and IPA for 10s. After the realization of PMMA nanostructures, 20 nm chromium (Cr) is deposited via E-beam evaporation and the nanostructures are transferred to Cr layer by a standard lift-off process. Taking the Cr nano-pattern as a hard mask, a slanted etching process of Si is performed using reactive ion etching (Oxford PlasmaPro 800). The tilt angle of nanostructure is controlled by the tilt stage and a Faraday cage. After removing Cr mask in chromium etchant solution for 5min, the tilted Si nano bars are obtained and then coated with 490 nm polyimide solution (PI-DMF) via a spin-coating process under negative pressure. The samples are solidified on a hotplate at 250°C for 1h and further coated with an antimony sulfide film via thermal evaporation. At last, a $SiO_2$ thin film is deposited as a protective layer using E-beam evaporation.

The phase transition process is all-optically controlled. For the amorphous to crystalline transition, a continuous wave (CW) laser at 450nm is focused onto the sample through an objective lens (20X, NA = 0.4). The size of focused beam is scanned on the sample to realize the phase transition of entire sample. For the transition from crystalline to amorphous, the CW laser is replaced by a frequency doubled femtosecond laser (100 fs pulse width, 1 kHz repetition rate, 400 nm wavelength).

### Optical characterization

Transmission spectra of metasurfaces are obtained with a standard setup. The circularly polarized incident beam is achieved by passing a supercontinuum laser (NKT) through a linear polarizer and a 1/4 waveplate. The incident light is focused by an objective lens (5X, NA = 0.12) on the sample, which is placed on a rotation translation stage. The transmitted beam is collected and collimated by the second objective lens (5X, NA = 0.12). The polarization components are analyzed by passing a linear polarizer and a 1/4 waveplate again and coupled to an infrared spectrometer (YOKOGAWA, AQ6370D). The angle-dependent transmission stage is achieved by rotating the sample with an angle resolution of 0.1°.

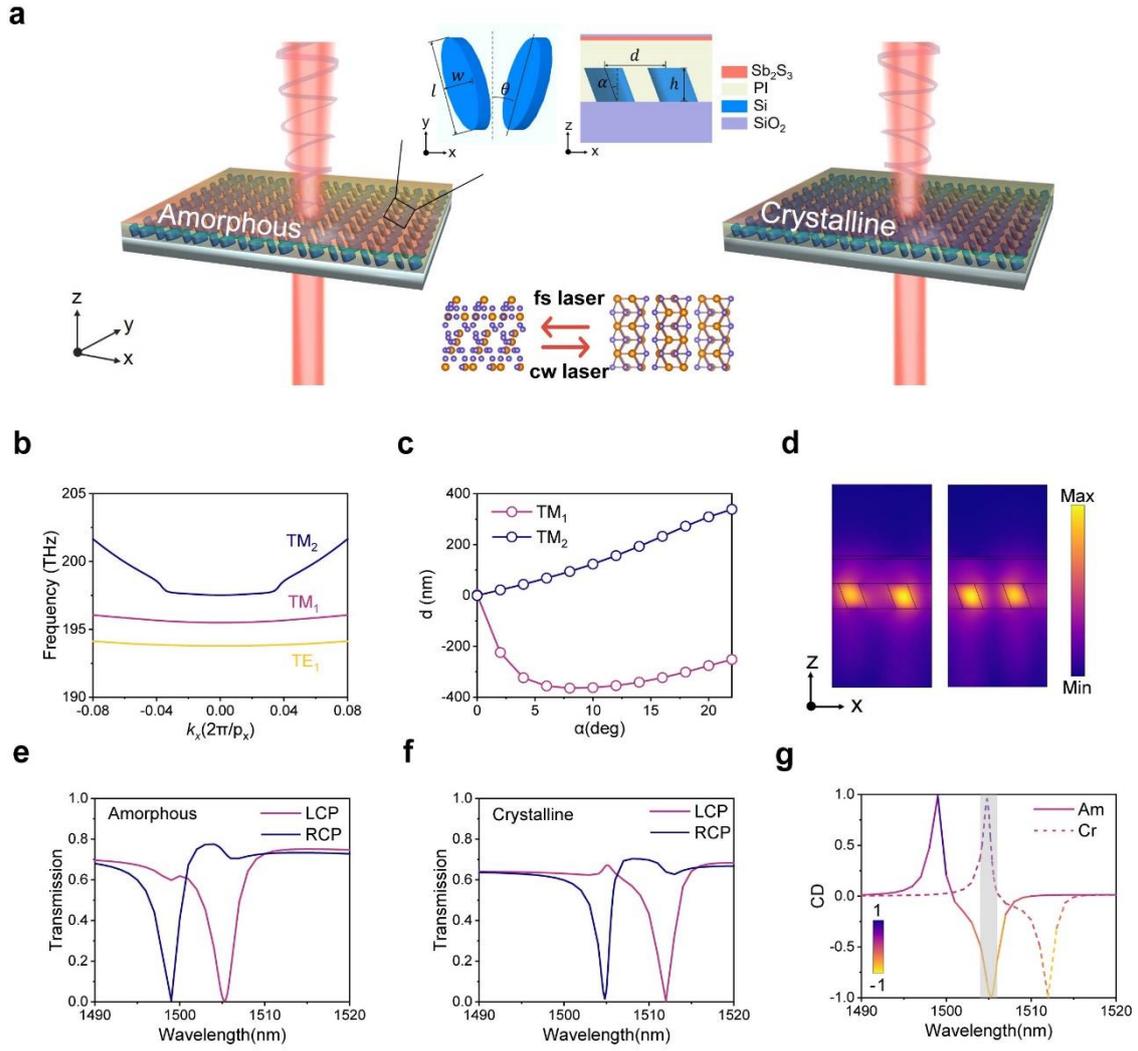

**Figure 1: Concept of tunable chiral metasurfaces.** (**a**) Schematic picture of hybrid phase-change metasurface. Insets show detailed structural parameters and transition between amorphous and crystalline phases of phase-change materials. The in-plane asymmetric parameter is $\theta = 15°$. (**b**) Band structure of phase-change metasurface at amorphous phase. Here the out-of-plane asymmetric parameter is $\alpha = 0°$. (**c**) Effective vertical offset of two modes as a function of out-of-plane asymmetric parameter $\alpha$. (**d**) Field distributions ($|H|$) of TM$_1$ mode (left) and TM$_2$ mode (right) in x-z plane. (**e**) and (**f**) are the LCP and RCP transmission spectra under amorphous and crystalline phase. (**g**) CD values correspond to (e) and (f). The sign of CD is reversed at 1505 nm (shadowed region).

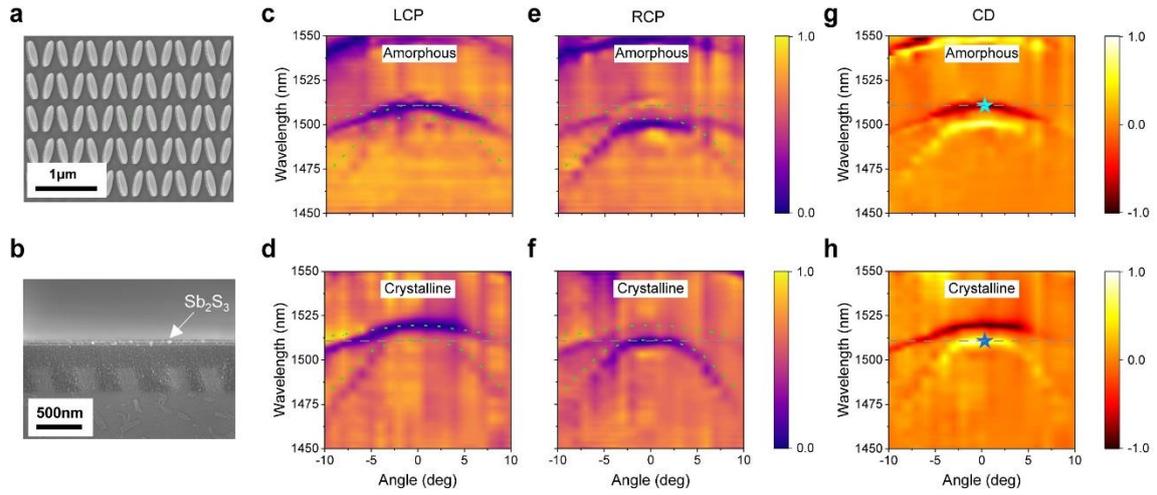

**Figure 2: Experimental characterization of chiral phase-change metasurfaces.** (**a**, **b**) Top-view and side view SEM images of the metasurface. Both the in-plane and out-of-plane asymmetries can be seen. (**c**, **e**) are the experimentally recorded LCP and RCP transmission spectra of metasurface under the amorphous phase. (**d**, **f**) Corresponding spectra under the crystalline phase. (**g**, **h**) Calculated CD values under amorphous phase and crystalline phase, respectively. The dashed lines in (c)-(f) represent the resonant wavelengths in numerical simulation.

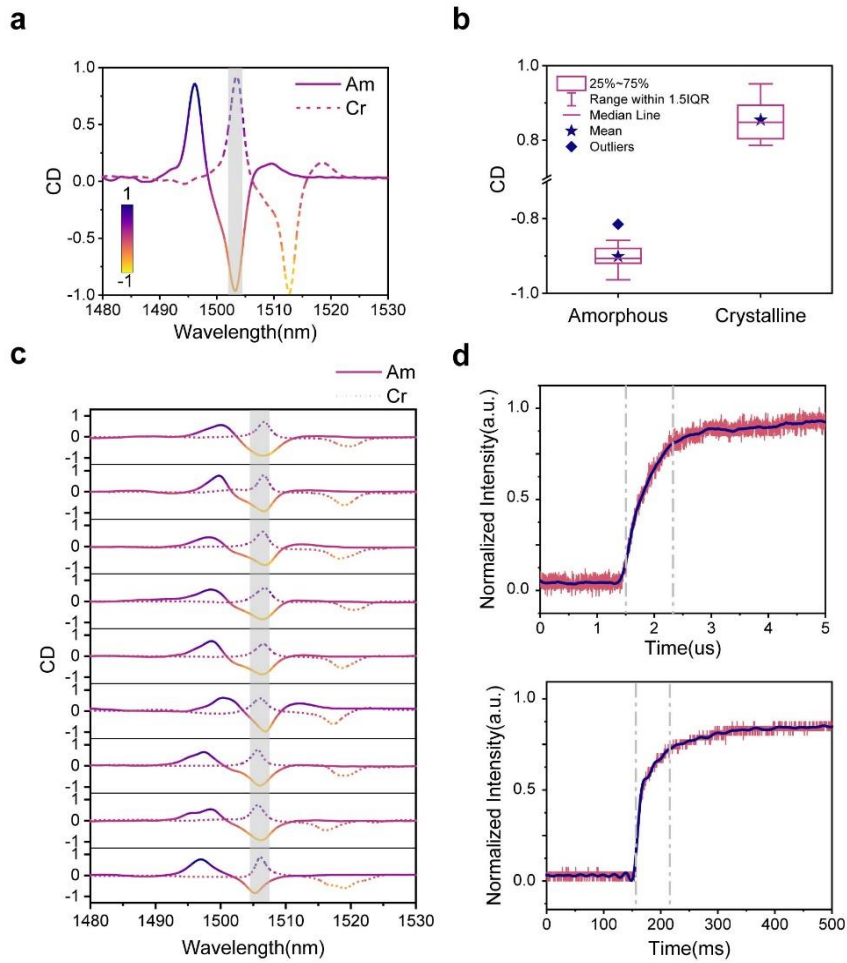

**Figure 3: Reversing chiral response** (**a**) The CD spectra of phase-change metasurface under amorphous and crystalline phase. (**b**) The averaged CD values of 12 phase-change metasurface at the same wavelength range under amorphous and crystalline phase. (**c**) The back-and-forth transition between amorphous and crystalline phase over multiple cycles for reversible chiral response. (**d**) are the transition time from amorphous to crystalline phase (bottom) and vice versa (top).

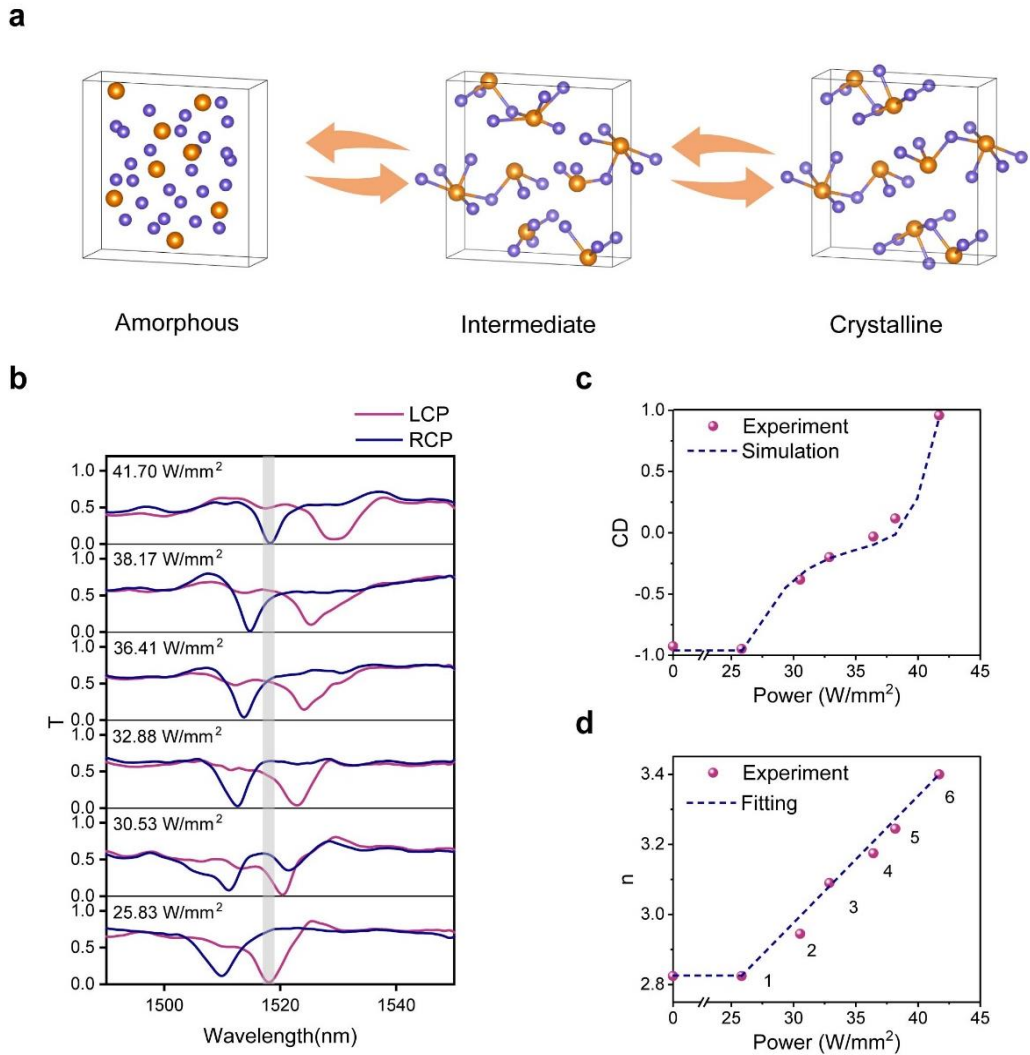

**Figure 4: Observation of chirality continuum.** (**a**) Transition from amorphous to intermediate and eventually to crystalline phase. (**b**) Transmission spectra of phase-change metasurface under different excitation power density. (**c**) Experimentally recorded CD values (dots) as a function of excitation power. (**d**) Retrieved refractive index of intermediate state (dots) from transmission spectra. The power-dependent refractive index (dashed line) is fitted by dots 1 and 6. Fitted refractive index is utilized to numerically simulate the CD values of phase-change metasurface at 1518 nm (dashed line in (c)).

**ACKNOWLEDGEMENTS**

The authors acknowledge support by National Key Research and Development Program of China (Grant Nos. 2021YFA1400802 and 2022YFA1404700), National Natural Science Foundation of China (Grant Nos. 6233000076, 12334016, 11934012, 12025402, 62125501, 12261131500 and 92250302), New Cornerstone Science Foundation through XPLORER PRIZE, Shenzhen Fundamental Research Project (Grant Nos. JCYJ20210324120402006, JCYJ20220818102218040, JCYJ20200109112805990 and GXWD20220817145518001) and Fundamental Research Funds for the Central Universities (Grant Nos. 2022FRRK030004, 2023FRFK03049).

**AUTHOR CONTRIBUTIONS**

Q.S., Y.K., S.X. conceived the idea and supervised the research. X.S., Y.Z. and J.Y. did the design. X.S., F.L. and H.Z. fabricated the samples. X.S. and K.D. performed the experimental measurements. Q.S., Y.K., S.X., J.H. and B.S. analyzed the results. All the authors discussed the contents and prepared the manuscript.

**COMPETING INTERESTS**

The authors declare no competing interests.